\documentclass{mem}

\usepackage{natbib}
\usepackage{txfonts}
\usepackage{balance}
\usepackage{graphicx}

\begin{document}

\title{Stratification in polarization and Faraday rotation in the jet of 3C~120}
\subtitle{}

\author{Jos\'e L. G\'omez\inst{1}, Iv\'an Agudo\inst{1}, Alan P. Marscher\inst{2}, Svetlana G. Jorstad\inst{2}  \and Mar Roca-Sogorb\inst{1}}

\institute{Instituto de Astrof\'{\i}sica de Andaluc\'{\i}a, CSIC, Apartado 3004, 18080 Granada, Spain. \email{jlgomez@iaa.es; iagudo@iaa.es; mroca@iaa.es} \and Institute for Astrophysical Research, Boston University, 725 Commonwealth Avenue, Boston, MA 02215, USA. \email{marscher@bu.edu; jorstad@bu.edu}}

\authorrunning{G\'omez et al.}
\titlerunning{Stratification in polarization and Faraday rotation in the jet of 3C~120}

\abstract{
  Very long baseline interferometric observations of the radio galaxy 3C~120 show a systematic presence of gradients in Faraday rotation and degree of polarization across and along the jet. These are revealed by the passage of multiple superluminal components throughout the jet as they move out from the core in a sequence of 12 monthly polarimetric observations taken with the VLBA at 15, 22, and 43 GHz. The degree of polarization has an asymmetric profile in which the northern side of the jet is more highly polarized. The Faraday rotation measure is also stratified across the jet width, with larger values for the southern side. Superposed on this structure we find a localized region of high Faraday rotation measure ($\sim$ 6000 rad m$^{-2}$) between approximately 3 and 4 mas from the core. This region of enhanced Faraday rotation may result from the interaction of the jet with the ambient medium, which may also explain the stratification in degree of polarization. The data are also consistent with a helical magnetic field in a two-fluid jet model, consisting of an inner emitting jet and a sheath of nonrelativistic electrons.


}
\maketitle{}

\section{Introduction}

  Helical magnetic fields may play an important role in the dynamics and emission of relativistic jets in active galactic nuclei (AGN), specially in the formation and collimation processes \citep{2002Sci...295.1688K,2008Natur.452..966M}. If jets are surrounded by a sheath of nonrelativistic electrons, it is possible to search for these helical magnetic fields by looking for Faraday rotation measure (RM) gradients across the jet \citep*[e.g.,][]{Blandford:1993fk}. Such gradients have been observed across the jet in 3C~273 \citep{2002PASJ...54L..39A,2008ApJ...675...79A,2005ApJ...626L..73Z,2005ApJ...633L..85A} and other sources \citep[e.g.,][]{2004MNRAS.351L..89G}, however they do not seem to be a universal feature \citep{2003ApJ...589..126Z}.

\begin{figure*}[t!]
\center{\includegraphics[scale=0.85]{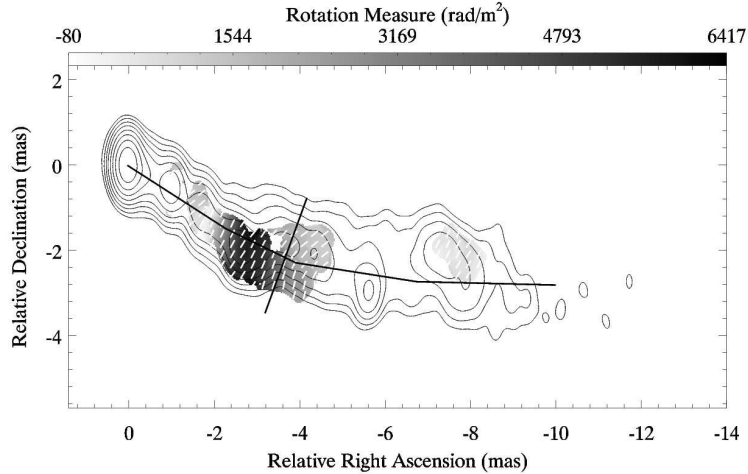}}
\caption{\footnotesize Map of the mean value of the rotation measure across epochs. Data with standard deviation larger than 1000 rad m$^{-2}$ were discarded. Bars indicate the mean value of the RM-corrected EVPAs, with all displayed pixels having a standard deviation smaller than 30$^{\circ}$ (96\% under 20$^{\circ}$). Contours show the 22 GHz total intensity at epoch 2001.00 for reference. The thick lines indicate the direction of the slices shown in Fig.~\ref{slices}.}
\label{rm_mean}
\end{figure*}

  Thanks to its proximity ($z=0.033$), Very Long Baseline Array (VLBA) observations of the radio galaxy 3C~120 are capable of resolving the jet across its width, revealing a very rich structure in total and polarized flux \citep{1998ApJ...499..221G,1999ApJ...521L..29G,2000Sci...289.2317G,2001ApJ...561L.161G,2008ApJ...681L..69G,2001ApJ...556..756W,2005AJ....130.1418J,2002Natur.417..625M,2007ApJ...665..232M}. In addition, evidence for the presence of a helical magnetic field has been found in 3C~120 by analyzing the motion and polarization of superluminal components \citep{2001ApJ...561L.161G,2005ApJ...620..646H}.

\section{Observations}

  The observations \citep{2008ApJ...681L..69G} were made with the 10 antennas of the VLBA at the standard frequencies of 15, 22, and 43 GHz, covering a total of 12 epochs: 2000 December 30, 2001 February 1, 2001 March 5, 2001 April 1, 2001 May 7, 2001 June 8, 2001 July 7, 2001 August 9, 2001 September 9, 2001 October 11, 2001 November 10, and 2001 December 13. The absolute phase offset between the right- and left-circularly polarized data, which determines the electric vector position angle (EVPA), was obtained by comparison of the integrated polarization of the VLBA images of several calibrators (0420$-$014, OJ~287, BL~Lac, and 3C~454.3) with VLA observations at epochs 2000 December 31, 2001 February 3, 2001 March 4, 2001 April 6, 2001 May 11, 2001 June 10, 2001 August 12, 2001 September 10, 2001 October 12, 2001 November 6, and 2001 December 15. Estimated errors in the orientation of the EVPAs usually lie in the range of 5$^{\circ}$-7$^{\circ}$. Comparison of the D-terms across epochs provides an alternative calibration of the EVPAs \citep{2002.VLBA.SM.30}, which was found to be consistent with that obtained by comparison with the VLA data.

\section{Faraday rotation and polarization gradients}

We have computed rotation measure maps at each epoch from the EVPA maps \citep{2008ApJ...681L..69G}, obtaining excellent fits to a $\lambda^2$ law of the EVPAs throughout the jet. Multiple superluminal components, with proper motions of the order of $\sim$ 2 mas yr$^{-1}$, sample the RM throughout the jet as they move out from the core. This allows derivation of the rotation measure image of Fig.~\ref{rm_mean} by computing the mean value of the rotation measure maps at each pixel. Figure \ref{rm_mean} reveals a localized region of enhanced rotation measure at a distance from the core of $\sim$ 3 mas, with a peak of $\sim$ 6000 rad m$^{-2}$.

\begin{figure}[t!]
\resizebox{\hsize}{!}{\includegraphics[clip=true]{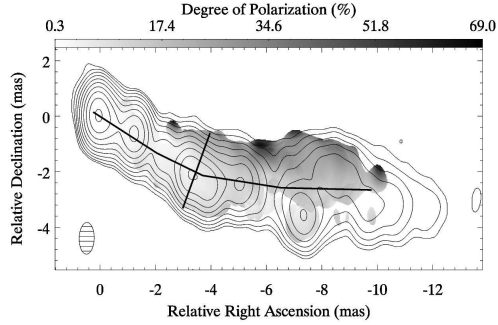}}
\caption{\footnotesize Map of the mean value of the degree of polarization at 15 GHz obtained by combining the data at all observed epochs. Total intensity contours (epoch 2001.86) are overlaid at 0.68, 1.4, 3.0, 6.3, 13, 28, 59, 125, 263, and 556 mJy beam$^{-1}$. The restoring beam of 1.14$\times$0.54 mas at position angle $-1.3^{\circ}$ is shown in the lower left corner. The thick lines indicate the direction of the slices shown in Fig.~\ref{slices}.}
\label{deg_pol}
\end{figure}

\begin{figure}[t!]
\resizebox{\hsize}{!}{\includegraphics[clip=true]{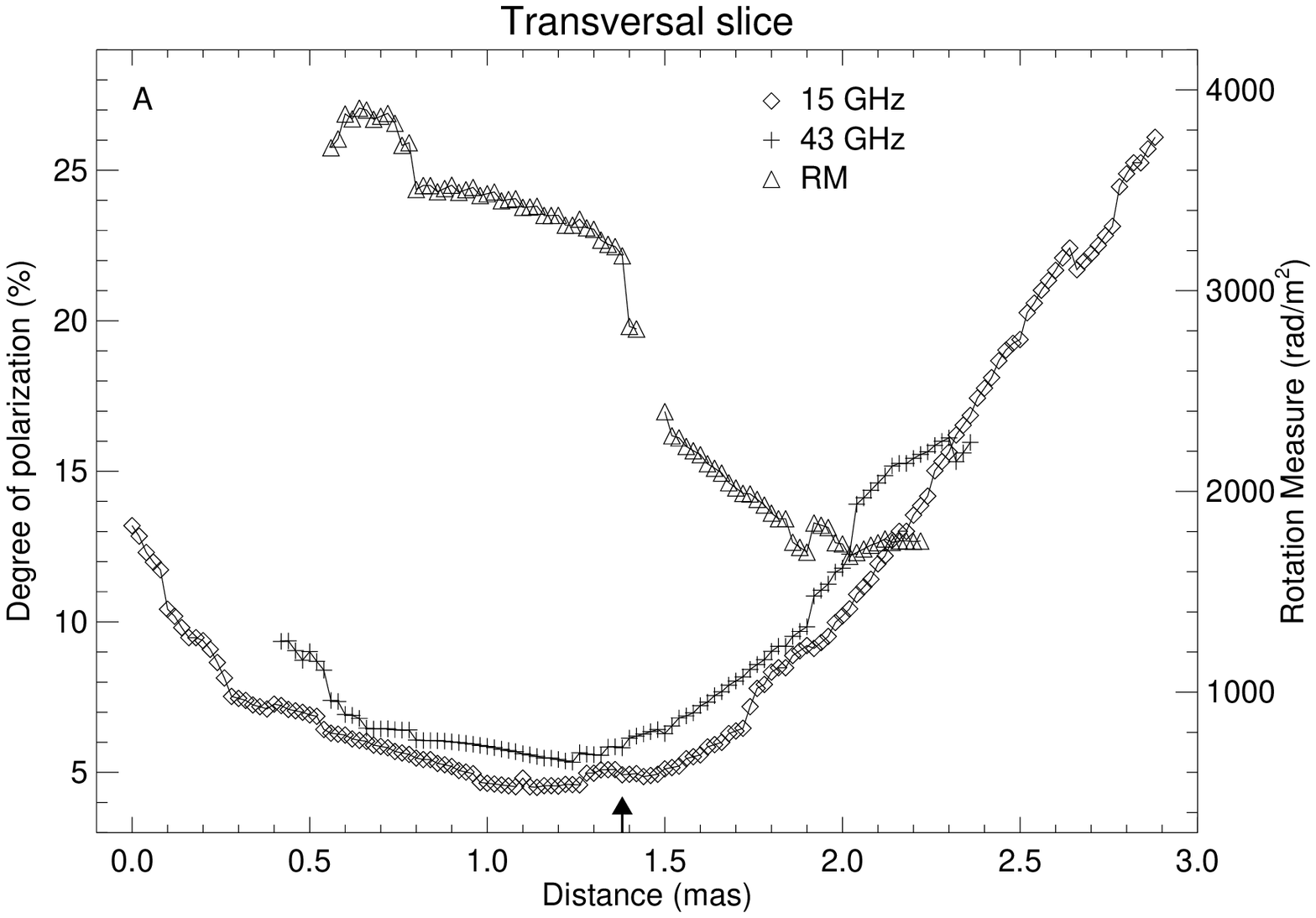}}
\resizebox{\hsize}{!}{\includegraphics[clip=true]{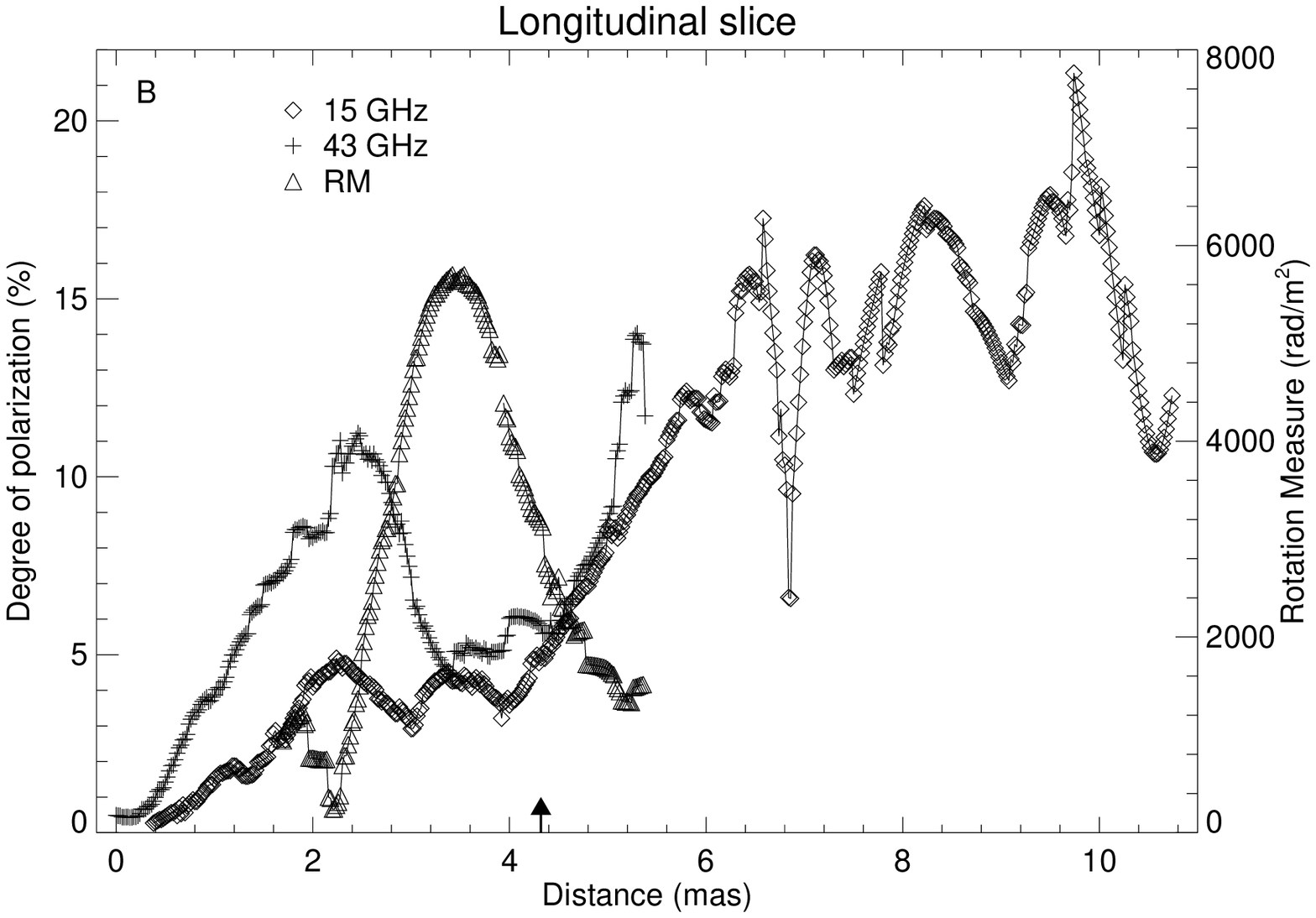}}
\caption{\footnotesize Degree of polarization and rotation measure slices across \emph{(a; top)} and along \emph{(b; bottom)} the jet, as shown by the thick lines of Figs.~\ref{rm_mean} and \ref{deg_pol}. The arrow indicates the location at which both slices intersect.}
\label{slices}
\end{figure}

  Figure \ref{deg_pol} shows the mean value of the degree of polarization (\emph{m}) at 15 GHz, revealing a clear stratification in polarization across the jet width, with significantly larger values on the northern side of the jet, also present at 22 and 43 GHz \citep{2008ApJ...681L..69G}. Figure \ref{slices} shows slices across and along the jet of the rotation measure and degree of polarization at 15 and 43 GHz. A transverse stratification in the rotation measure appears in Fig.\ref{slices}a, with larger values on the southern side of the jet.

  The transverse profile of \emph{m} suggests a progressive reordering of the magnetic field (as integrated along the line of sight) toward the jet edges. This is consistent with the presence of a helical magnetic field and/or a shear layer produced by the interaction of the jet with the external medium \citep{2000ApJ...528L..85A,2005MNRAS.360..869L}. A helical magnetic field could also explain the observed asymmetry in the \emph{m} profile, and is in agreement with the observed transverse profile of RM.

  By comparing the transverse profiles of \emph{m} and RM (Fig.~\ref{slices}a) we find that there is not a strong dependence of \emph{m} on RM and frequency, as expected for the case of internal Faraday rotation \citep{1966MNRAS.133...67B}. Hence, although we cannot rule out some internal Faraday depolarization, the transverse profiles of \emph{m} and RM are more consistent with an external RM screen. In this case a two-fluid model, with an internal emitting jet and a sheath of thermal electrons, both immersed in a helical magnetic field, could provide an interpretation for the observed transverse profiles of \emph{m} and RM. The RM-corrected EVPAs, predominantly perpendicular to the jet axis (Fig.~\ref{rm_mean}), require a dominant poloidal (as measured in the frame of the jet fluid) magnetic field in the emitting region \citep{2005MNRAS.360..869L}.

  The longitudinal profile (Fig.~\ref{slices}b) shows a progressive increase in \emph{m} with distance along the jet, from an unpolarized core to 15-20\% beyond $\sim$8 mas. The low polarization and its strong dependence on frequency in the inner jet regions ($\lesssim$3 mas) may be related to opacity effects, or caused by a very high and time/space variable RM \citep*[G\'omez et al., in preparation; see also][]{2005ApJ...633L..85A,2007AJ.134.799J}. A progressive reordering of the magnetic field may be required to explain the subsequent increase in degree of polarization.

\section{Interaction with the ambient medium}

  Figure \ref{slices}b reveals the longitudinal profile of the localized region of large RM shown in Fig.~\ref{rm_mean}. The degree of polarization has a clear dependence with RM, suggesting that the decay is produced by Faraday depolarization. The rapid decrease in \emph{m} implies that there may be some internal Faraday depolarization, but it is difficult to test without further information regarding the intrinsic values of \emph{m}. This region is coincident in location, and has similar values of the RM to those postulated by \citet{2000Sci...289.2317G} from two-frequency observations. A local process, such as interaction of the jet with the external medium or a cloud, would be required in order to explain the existence of this region, as previously suggested by \citet{2000Sci...289.2317G}. The observed Faraday rotation can originate from an ionized cloud along the line of sight that may also physically interact with the jet. Some internal Faraday depolarization could also be expected if the jet partially entrains some of the thermal material of the external medium/cloud.

  Our observations therefore support the conclusion that interaction of the jet with the external medium causes the excess Faraday rotation, stratification in the degree of polarization across the jet, and flux density flares of superluminal knots a few mas from the core. A helical magnetic field in a two-fluid jet model can be accommodated within this scenario, but by itself cannot explain the existence of the localized Faraday rotation region.

\acknowledgements This research has been supported by the Spanish Ministerio de Educaci\'on y Ciencia and the European Fund for Regional Development through grants AYA2004-08067-C03-03 and AYA2007-67627-C03-03, and by National Science Foundation grant AST-0406865. I.\ A. acknowledges support by an I3P contract by the Spanish Consejo Superior de Investigaciones Cient\'{\i}ficas. The VLBA is an instrument of the National Radio Astronomy Observatory, a facility of the National Science Foundation operated under cooperative agreement by Associated Universities, Inc.


\begin{thebibliography}{}

\bibitem[\protect\citeauthoryear{Aloy et~al.}{Aloy
  et~al.}{2000}]{2000ApJ...528L..85A}
Aloy M.~A., G{\'o}mez J.~L., Ib{\'a}{\~n}ez J.~M., Mart{\'\i} J.~M.,
  M{\"u}ller E., 2000, ApJ, 528, L85

\bibitem[\protect\citeauthoryear{Asada et~al.}{Asada
  et~al.}{2008}]{2008ApJ...675...79A}
Asada K., Inoue M., Kameno S.,  Nagai H., 2008, ApJ, 675, 79

\bibitem[\protect\citeauthoryear{Asada et~al.}{Asada
  et~al.}{2002}]{2002PASJ...54L..39A}
Asada K., Inoue M., Uchida Y., et~al., 2002, PASJ, 54, L39

\bibitem[\protect\citeauthoryear{Attridge, Wardle, \& Homan}{Attridge
  et~al.}{2005}]{2005ApJ...633L..85A}
Attridge J.~M., Wardle J.~F.~C.,  Homan D.~C., 2005, ApJ, 633, L85

\bibitem[\protect\citeauthoryear{Blandford}{Blandford}{1993}]{Blandford:1993fk}
Blandford R.~D., 1993, in D.~Burgarella C.~P.~O., M.~Livio (ed.), Astrophysical
  Jets.
\newblock Cambridge: Cambridge Univ. Press, p.~15

\bibitem[\protect\citeauthoryear{Burn}{Burn}{1966}]{1966MNRAS.133...67B}
Burn B.~J., 1966, MNRAS, 133, 67

\bibitem[\protect\citeauthoryear{Gabuzda, Murray, \& Cronin}{Gabuzda
  et~al.}{2004}]{2004MNRAS.351L..89G}
Gabuzda D.~C., Murray {\'E}.,  Cronin P., 2004, MNRAS, 351, L89

\bibitem[\protect\citeauthoryear{G{\'o}mez, Marscher, \& Alberdi}{G{\'o}mez
  et~al.}{1999}]{1999ApJ...521L..29G}
G{\'o}mez J.~L., Marscher A.~P.,  Alberdi A., 1999, ApJ, 521, L29

\bibitem[\protect\citeauthoryear{G{\'o}mez et~al.}{G{\'o}mez
  et~al.}{2001}]{2001ApJ...561L.161G}
G{\'o}mez J.~L., Marscher A.~P., Alberdi A., Jorstad S.~G.,  Agudo I., 2001,
  ApJ, 561, L161

\bibitem[\protect\citeauthoryear{G{\'o}mez et~al.}{G{\'o}mez
  et~al.}{2002}]{2002.VLBA.SM.30}
G{\'o}mez J.~L., Marscher A.~P., Alberdi A., Jorstad S.~G.,  Agudo I., 2002,
  VLBA Scientific Memo

\bibitem[\protect\citeauthoryear{G{\'o}mez et~al.}{G{\'o}mez
  et~al.}{2000}]{2000Sci...289.2317G}
G{\'o}mez J.~L., Marscher A.~P., Alberdi A., Jorstad S.~G.,
  Garc{\'\i}a-Mir{\'o} C., 2000, Science, 289, 2317

\bibitem[\protect\citeauthoryear{G{\'o}mez et~al.}{G{\'o}mez
  et~al.}{1998}]{1998ApJ...499..221G}
G{\'o}mez J.~L., Marscher A.~P., Alberdi A., Mart{\'\i} J.~M.,  Ib{\'a}{\~n}ez
  J.~M., 1998, ApJ, 499, 221

\bibitem[\protect\citeauthoryear{G{\'o}mez et~al.}{G{\'o}mez
  et~al.}{2008}]{2008ApJ...681L..69G}
G{\'o}mez J.~L., Marscher A.~P., Jorstad S.~G., Agudo I.,  Roca-Sogorb M.,
  2008, ApJ, 681, L69

\bibitem[\protect\citeauthoryear{Hardee, Walker, \& G{\'o}mez}{Hardee
  et~al.}{2005}]{2005ApJ...620..646H}
Hardee P.~E., Walker R.~C.,  G{\'o}mez J.~L., 2005, ApJ, 620, 646

\bibitem[\protect\citeauthoryear{Jorstad et~al.}{Jorstad
  et~al.}{2005}]{2005AJ....130.1418J}
Jorstad S.~G., Marscher A.~P., Lister M.~L., et~al., 2005, AJ, 130, 1418

\bibitem[\protect\citeauthoryear{Jorstad et~al.}{Jorstad
  et~al.}{2007}]{2007AJ.134.799J}
Jorstad S.~G., Marscher A.~P., Stevens J.~A., et~al., 2007, AJ, 134, 799

\bibitem[\protect\citeauthoryear{Koide et~al.}{Koide
  et~al.}{2002}]{2002Sci...295.1688K}
Koide S., Shibata K., Kudoh T.,  Meier D.~L., 2002, Science, 295, 1688

\bibitem[\protect\citeauthoryear{Lyutikov, Pariev, \& Gabuzda}{Lyutikov
  et~al.}{2005}]{2005MNRAS.360..869L}
Lyutikov M., Pariev V.~I.,  Gabuzda D.~C., 2005, MNRAS, 360, 869

\bibitem[\protect\citeauthoryear{Marscher et~al.}{Marscher
  et~al.}{2008}]{2008Natur.452..966M}
Marscher A.~P., Jorstad S.~G., D'Arcangelo F.~D., et~al., 2008, Nature, 452,
  966

\bibitem[\protect\citeauthoryear{Marscher et~al.}{Marscher
  et~al.}{2002}]{2002Natur.417..625M}
Marscher A.~P., Jorstad S.~G., G{\'o}mez J.~L., et~al., 2002, Nature, 417, 625

\bibitem[\protect\citeauthoryear{Marscher et~al.}{Marscher
  et~al.}{2007}]{2007ApJ...665..232M}
Marscher A.~P., Jorstad S.~G., G{\'o}mez J.~L., et~al., 2007, ApJ, 665, 232

\bibitem[\protect\citeauthoryear{Walker et~al.}{Walker
  et~al.}{2001}]{2001ApJ...556..756W}
Walker R.~C., Benson J.~M., Unwin S.~C., et~al., 2001, ApJ, 556, 756

\bibitem[\protect\citeauthoryear{Zavala \& Taylor}{Zavala \&
  Taylor}{2003}]{2003ApJ...589..126Z}
Zavala R.~T.,  Taylor G.~B., 2003, ApJ, 589, 126

\bibitem[\protect\citeauthoryear{Zavala \& Taylor}{Zavala \&
  Taylor}{2005}]{2005ApJ...626L..73Z}
Zavala R.~T.,  Taylor G.~B., 2005, ApJ, 626, L73

\end{thebibliography}
\bibliographystyle{aa}

\end{document}